\documentclass{article}
\usepackage[T1]{fontenc} 
\usepackage[utf8]{inputenc} 
\usepackage{ismir,amsmath,cite,url}
\usepackage{graphicx}
\usepackage{color}
\usepackage{amssymb}
\usepackage{xspace}
\usepackage{cleveref}

\newcommand{\hooknet}{Hook-Net\xspace}

\newcommand{\mfull}{\hooknet}
\newcommand{\esclick}{SigClick\xspace}

\newcommand{\mpdrei}{MP3\xspace}
\newcommand{\Mpdrei}{MP3\xspace}

\title{Audio Defect Detection in Music with Deep Networks}

\oneauthor
{Daniel Wolff, Rémi Mignot and Axel Roebel}
{Analysis-Synthesis Team, UMR 9912 STMS - IRCAM, CNRS, Sorbonne Université, Paris, France\\
\tt{\{daniel.wolff, remi.mignot, axel.roebel\}@ircam.fr}}

\begin{document}

\maketitle

\begin{abstract}
	With increasing amounts of music being digitally transferred from production to distribution, automatic means of determining media quality are needed. 
	Protection mechanisms in digital audio processing tools have not eliminated the need of production entities located downstream the distribution chain to assess audio quality and detect defects inserted further upstream.
	Such analysis often relies on the received audio and scarce meta-data alone.
	Deliberate use of artefacts such as clicks in popular music as well as more recent defects stemming from corruption in modern audio encodings 
	call for data-centric and context-sensitive solutions for detection. 
	We present a convolutional network architecture following end-to-end encoder-decoder configuration to develop detectors for two exemplary audio defects. 
	A click detector is trained and compared to a traditional signal processing method, with a discussion on context sensitivity. 
	Additional post-processing is used for data augmentation and workflow simulation.
	The ability of our models to capture variance is explored in a detector for artefacts from decompression of corrupted \mpdrei compressed audio.
	For both tasks we describe the synthetic generation of artefacts for controlled detector training and evaluation.
	We evaluate our detectors on the large open-source Free Music Archive (FMA) and genre-specific datasets.
\end{abstract}

\section{Introduction}\label{sec:introduction}
In recent decades, digital means of media delivery have become increasingly popular, not least with the advent of high-speed internet and the ubiquity of digital playback and recording devices ranging from studio digital signal processing hardware  to \mpdrei players and mobile phones.	
The greater availability of technology required to produce digital media now allows for small-scale studios to create high quality content and instantly transfer it to distributors such as music labels.

Digital media files can suffer from various degradations that occur during transport and processing of the media.
Automatic means exist for detection and correction of certain data errors in uncompressed audio, but many complex defects remain untackled.
In this paper, we explore the efficacy of applying a data-driven machine learning approach using Deep Neural Networks to two audio defects, which music labels are confronted with in quality assurance of incoming media.

For our first scenario, we define clicks as discontinuities affecting a few signal samples, resulting in very short broadband impulses. 
A prominent source of similar artefacts are buffer under-runs, where, due to synchronisation issues during digital audio processing, a few samples of an old or zeroed signal are transmitted instead of the current signal.
Although existing mastering software offers methods to remedy similar artefacts such as clicks, crackle and clipping, 
restoration often requires manual selection of noise profile, target segments or thresholds. 
This is due to ambiguities introduced by e.g. signal quality or, depending on the genre, degradations voluntarily applied to audio as effects, rendering current methods costly in large-scale application. 

Our hypothesis is that, using a data-driven approach, our network can distinguish deliberate clicks (such as electronic snare drums) from defects, and thereby enable automatic processing of large electronic music corpora.
This is an essential challenge in our scenario, where manual inspection may not be feasible because of operational constraints.

In our second scenario, we aim to detect corruption of binary \mpdrei\footnote{See standards  \url{https://www.iso.org/standard/19180.html} and  \url{https://www.iso.org/standard/31537.html}} data via the artefacts audible after decoding such files.
The \mpdrei audio encoding family uses a psycho\-acoustic model to guide data reduction.
Due to the transformations applied to data during \mpdrei decoding, a large variation of effects is possible, ranging from added whistling noise over various missing frequency bands to broadband noise.
In contrast to noise normally added during the process of lossy encoding itself, the artefacts stemming from unnoticed corruption have not been approached for detection yet. 

Such \mpdrei corruption may happen within a production chain where manual transcoding is performed by different production agents, as illustrated with the following use case: 
A distributor receives a sound file in lossless format (e.g. pulse-code modulation format \texttt{.wav}) as a studio quality delivery from a production entity.
A degradation is detected at this point by listening to the audio.
During discussion of this defect, it is found that the production entity sent a transcoded file they prepared for checking an earlier, defect \mpdrei delivery.
The decoded \mpdrei was transmitted rather than the original \texttt{.wav} file intended for the delivery. 
Having no dependable meta-data on intermediary processing, the distributor can only rely on the audio itself for quality assurance. 
We found the above case to have practical relevance in the music industry, and designed our method to aid error detection in similar circumstances.

Our two detection scenarios are chosen to benefit from our data-driven approach: while the click detection may make use of signal context to determine the ``musicality'' of a click candidate, the \mpdrei glitch detector can benefit from the network's ability to capture the variation of artefacts during training.
In contrast to the click artefacts, the artistic use of the \mpdrei glitch artefacts is currently limited, reducing the chance to confuse such intentional use as degradation. 
The resulting models are designed for robust scanning of large media libraries in an unsupervised batch processing scenario.

In the following we present: an adaptation of the Wave-U-Net deep architecture to the detection and localisation of audio defects, two separate detectors built upon this to respectively detect clicks and \mpdrei glitches, methods for simulation of these artefacts assuring significance of inserted defects, and a large-scale evaluation against several music datasets.

\section{Related Work}\label{sec:relwork}
Traditional methods for detecting clicks and similar non-stationary noise in audio aim at detecting discontinuities, using autoregressive modelling in the raw audio or sparse optimisation in the spectral domain\cite{godsill2006digitalaudiorestauration, adler2011declipping}. 
For restoration of analog media, where issues like clicks, overload and high frequency noise are common, wavelet-based implementations are used in commercial noise removal software \cite{berger1994removing, laney2011}.
Recently, Deep Neural Networks have been used for noise and obstruction removal mostly in images but also for audio and other time-based signals\cite{Valin18SpeechNoise,MaasSpeechNoise12,ChiangECGNoise19}. 
Removal of obstructions from images is a task particularly close to our task as it deals with less stationary noise \cite{Xiang2019Deraining,Wang18Image2Image}. 
Matsui et al.\cite{Matsui2020UNetFence} use a convolutional neural network similar to our architecture to remove fences from images.
Restoration methods differ from our detection task in that explicit detection reporting and evaluation of false positives are not needed when applied as an audio de-noising effect.  

For lossy audio encodings such as \mpdrei, the artefacts arising during encoding, mainly as a trade-off between quality and bitrate, vary with methods and encoders and have been extensively discussed for the \mpdrei format \cite{brandenburg1999mp3}. 
Nevertheless, corruption of files introduces new, different artefacts that may remain unnoticed during decoding and thus are the subject of our detector.

Research in the related field of Computational Auditory Scene Analysis (CASA) concerns itself with the detection and labelling of events in an audio stream.
Deep Neural Networks are now being increasingly used in this field, often with a spectral feature extraction pre-processing step.  
Mesaros et al.\cite{MesarosASAClick2014} report the detection of ``clicks'' as one of 61 classes in a detection task on their private dataset, with a recognition rate of around 65 percent. 
It is not clear how these clicks compare to the clicks stemming from digitally signal failures which we tackle in this paper. 
The focus in CASA is to detect recorded sounds while being robust to noise in the recording, thus clicks to be detected would relate to physical events (see "mouse click" in the CASA dataset \cite{Giannoulis13adatabase}).
From a CASA perspective, intentional clicks, as frequently found in electronic music, would not necessarily be distinguished from those stemming from defects. 

This difficulty of ambiguity in audio defect annotation is noted by Alonso-Jiménez et al.\cite{alonso:2019:audiodefects} who describe their implementation and evaluation of established audio-defect detection algorithms.
Their optimised algorithms detect a significant number of audio defects in a database from a commercial streaming service, noting that there may be a long tail of likely, but undetected degradations. 
In our experiments (\Cref{sec:experimentsclick}) we evaluate their click detector implementation after Vaseghi\cite{Vaseghi06}, complementing their results with  accuracy metrics on a large dataset containing synthesized defects.

Research on error concealing scenarios, where defects are already identified, e.g. due to missing packets at the network layer during transport of an audio stream\cite{perkins2010packetloss}, can inform us of the complexity and artefacts resulting from such failures. 
Deep convolutional networks have been recently introduced into the field of audio inpainting - filling a gap in audio in such a way that the error is concealed - to great effect\cite{lee2015packetloss,marafioti2019audioinpainting}.
Their ability to encode and model variation in large datasets results in more intelligible speech reconstruction, when compared to conventional concealment approaches.

We aim to exploit gains from training with large datasets for our detectors.
Although large datasets of music, such as the FMA dataset, are available, we are not aware of any open datasets with audio defect labels. 
This may be due to the fact that defects are mainly corrected in production, and are - in terms of playback time - very rare.
The task of audio anomaly detection deals with this issue using unsupervised learning to model usual/common signals on unlabeled data. 
Autoregressive networks\cite{Rushe2019autoreganomaly} and autoencoders\cite{Marchi2016AutoEncNovelty} have recently been used to detect unusual acoustic events via their high reconstruction loss after such training.
Our architecture is similar to the above, in that it shares an information bottleneck to learn representations, but we use a supervised learning approach with synthetic examples similar to the work in \cite{Muehlbauer2010AudioDefects} to better control the type of artefacts detected, avoiding the detection of e.g. new instruments or audio samples as anomalies. 

Ronneberger et al. \cite{ronneberger2017unet} originally presented the U-Net as a deep convolutional neural network for the task of segmenting biomedical images. 
The network structure allows for efficient learning of spatial, or - in the case of audio - temporal and frequency patterns.
It has since been used for various end-to-end audio transformation tasks \cite{jansson2017unet}.
Stoller et al. \cite{stollerewert2018waveunet} modified this structure to work directly on the one-dimensional audio signal as input, resulting in the Wave-U-Net.
In their source-separation task, they employ the network on overlapping excerpts of the original signal at a low 8kHz sample rate.

\section{Model Architecture}\label{sec:flatunet}
In this paper we introduce the \hooknet as a novel adaptation of the Wave-U-Net for the task of detecting artefacts in audio signals.
We apply this model as an end-to-end approach, feeding raw waveform segments into the network to receive a time series of classification results.
After initial experiments with a U-Net on spectrogram features, we found training to be more effective when using raw audio input, which may be due to the extreme brevity of our defects.
The time-aligning horizontal connections of input and output promise to help capture the context of distortions in the input waveform. 

Our network takes as input segments 16384 samples of audio (at 44100Hz sample rate) and outputs a time series of 128 output samples that are individually quantised to the binary decision on original vs. degraded. 
The original Wave-U-Net implies that input and output share the same sampling rate.
To reduce computational cost while operating at source input sample rates, the \hooknet introduces an imbalance: the output time resolution is reduced by a factor of 128 with regards to the input sample rate, resulting in a classification sample rate of 344.5Hz.\footnote{This also allows comparisons with spectrogram-based models, which did not perform as well and are omitted for brevity.} 

\begin{figure}[htb]
	\centerline{\includegraphics[width=\columnwidth]{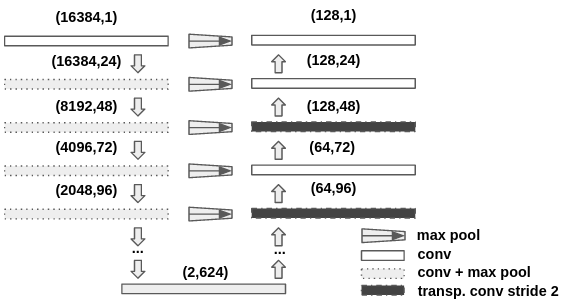}}
	\vspace{-0.3cm}
	\caption{Architecture of the \hooknet }
	\label{fig:uwunetarch}
\end{figure}
This is reflected in the network structure as displayed in \Cref{fig:uwunetarch}. 
Our contracting (left side) path consists in blocks comprising two sequences of (zero-padded convolution -- batch-norm -- activation) layers followed by max-pooling.
Here, the temporal resolution is halved every block, while the number of filters increases.
An expanding block consists of an upscaling operation, followed by concatenation of the skip connection from the contracting path and a regular convolution layer. 
In our model, resolution of the axis mapped to time in our output is only doubled every second block - resulting in a reduced temporal resolution at the network output.

For the upscaling operation we follow the original U-Net \cite{ronneberger2017unet}, but employ transposed convolution of stride 2 only every second layer, and stride 1 otherwise. 
Thereby some steadiness remains in the growth of resolution across the expanding path, despite the reduced final output resolution.
For the horizontal skip-connections, connecting the contracting and expansive path, we use max-pooling to adapt the time resolutions between the corresponding layers.
We furthermore add vertical skip connections on the contracting path, bridging every block of two convolutions. 
This strategy is motivated by the training benefits reported in residual networks\cite{kaiming2015deepresiduallearning}. 

In the following we describe the generation of click and \mpdrei glitch artefacts. 
	
\section{Click artefact generation}\label{sec:clicking}
Within the scope of this paper, a click degradation corresponds to a fault in the already digitised signal:  
we define a click as a discontinuity, where the signal changes sharply to a random value for 1-3 samples but then continues unchanged.
With this definition we aim to cover and simulate defects from digital signal transport, commonly resulting from buffer under-run during playback or mixing in a DAW or timing errors during digital transport over wire.

Clicks are inserted on-the-fly into the network input audio segments for training, validation and test scenarios. 
The position of the click is randomised, and one click is inserted with a probability of 0.1 per audio segment, with a small variation in length.
The amplitude value of the click is calculated as a random offset of the current signal, from a uniform distribution within $[0.3,1)$.
Each initial offset sign is chosen randomly, then signs that would create clipping are inverted.

The minimum amplitude offset ($0.3$) of the inserted clicks' amplitudes is introduced to assure that the signal change and resulting degradation is significant. 
In absence of perceptual data on the acoustic salience of inserted artefacts, this heuristic should create clicks that are likely to be audible, where they are not perceptually masked by close preceding transients or loud broad-band noise.

\subsection{Audio post-processing via SoX}\label{sec:augmentation}
In order to simulate potential post-processing effects that may have occurred on audio signals with previously undetected clicks, we use the SoX\footnote{\url{http://sox.sourceforge.net/}} sound processor to slightly alter the audio segments after the click-insertion steps.
Segments are post-processed regardless of whether a click has actually been inserted.
A random combination of reverb, two-band EQ and compression is applied, and strength and filter parameters are chosen randomly within ranges that apply only mild changes to the signal.
In \Cref{sec:experimentsclick} we report results for click detector training and detection with and without post-processing. 

\subsection{Click target vector}
The target vector for training and testing of the detector is a 128-component floating point vector that is set to 1 on the (resampled) position corresponding to the location of an inserted click, and 0 otherwise.
In our experiments we simulate the problem of rare clicks that may be overlooked during production.
There is at most 1 click per input segment. 

\section{\Mpdrei glitch artefact  generation}\label{sec:glitching}
This use-case tackles degradations that result from data corruption in the commonly used  MPEG-1/2 Audio Layer III (\mpdrei) lossy audio compression format. 
We will refer to these as \emph{glitch} defects.
This degradation is interesting as it can easily be ``overlooked'' in quality assurance. 
Moreover, the generation approach described below can be generalised to other audio codecs.

\begin{figure}[htb]
	\centerline{\includegraphics[trim=0 20 0 95,clip, width=\columnwidth]{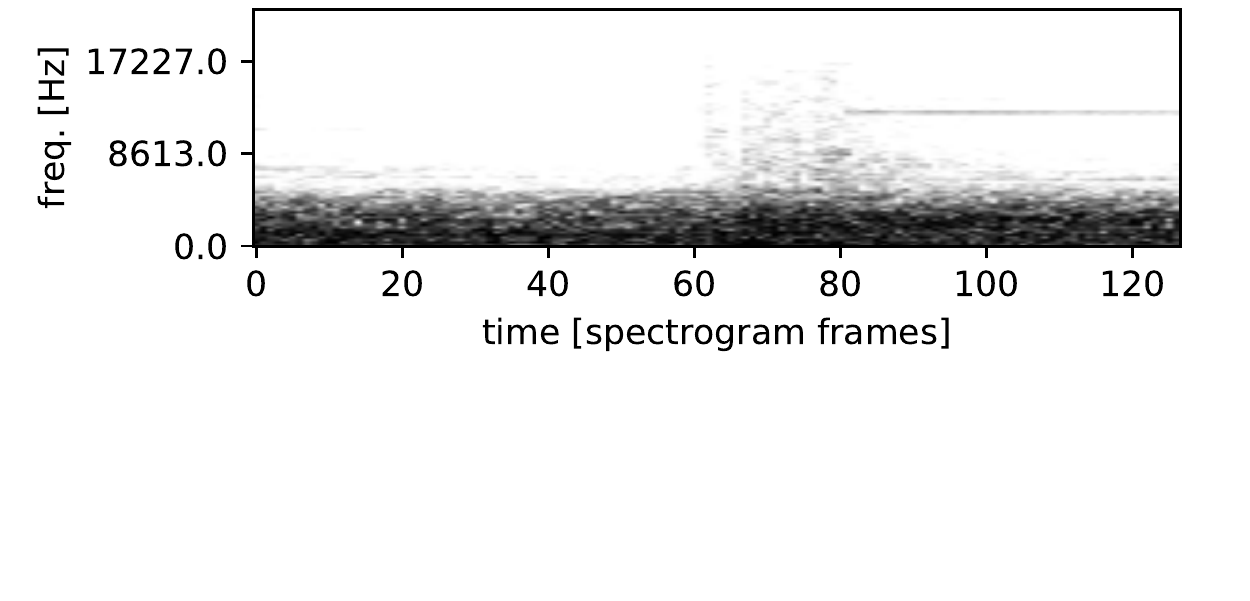}}
	\vspace{-2cm}
	\centerline{
		\includegraphics[trim=0 0 0 95,clip,width=\columnwidth]{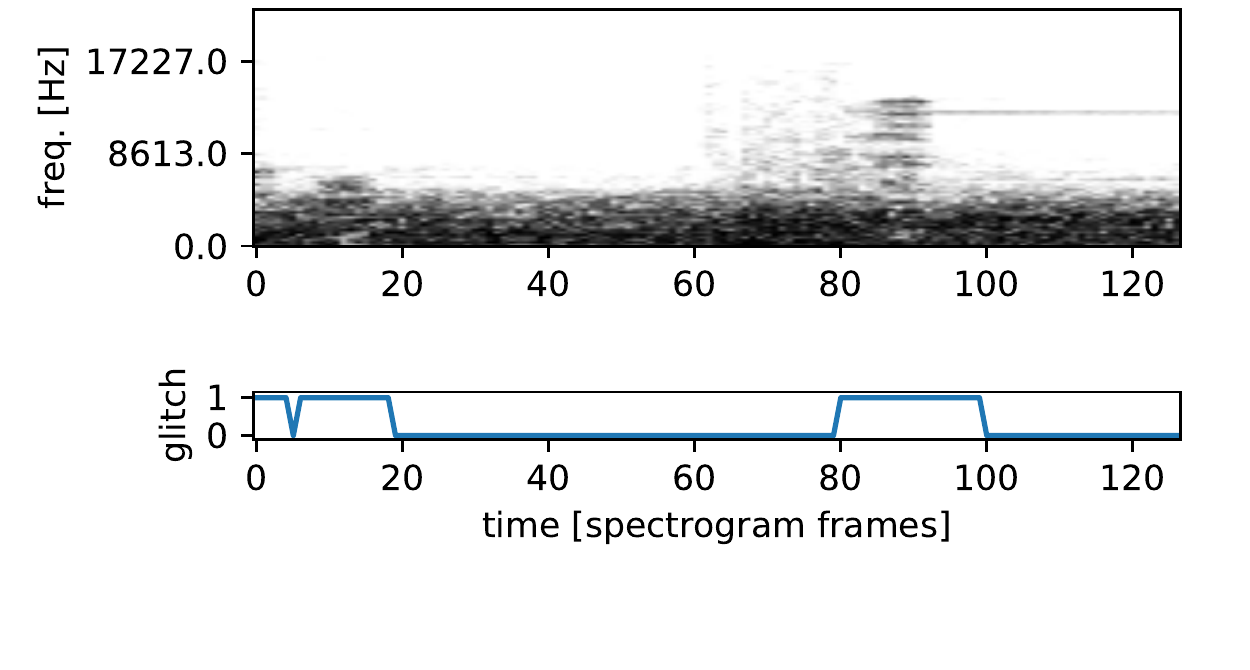}}
	\vspace{-1.0cm}
	\caption{Typical example of glitch degradation. Top: linear frequency spectrogram of original signal. Bottom: same for signal with two degradations at frame 0 (intermittent) and frame 80 with binary ground truth annotations from spectrogram comparison.
	}
	\label{fig:glitch_hf}
\end{figure}

The acoustic shape of the degradation varies strongly and often depends on the content and amplitude of the surrounding audio data.
Fig.\ref{fig:glitch_hf} displays a typical glitch artefact: the degradation consists in added high-frequency content. 
This effect appears when comparing the original (top) and degraded (bottom) spectrogram contents between 80-100 frames.
The same example features another glitch effect at the start of the frame, affecting a lower part of the spectrum.
Acoustically, the effect is similar to a short whistle. 
An audio example demonstrating glitch artefacts is available online\footnote{\url{https://osf.io/uqner/?view_only=042774933537440299dd48a4083305b1}}.
Note that the overall signal energy does not always change.
The often well-embedded and adaptive nature of these glitches render automatic detection difficult.

\subsection{Simulation of data corruption}\label{sec:glitchprod}
Unlike the click degradations, which can be inserted on-the-fly as in \Cref{sec:clicking}, \mpdrei glitch degradations are calculated on a per-piece basis before training due to the less easily indexable file format of \mpdrei which hinders exact seeking.

In order to simulate data corruption in the compressed format, we modify \mpdrei encoded data during a decoding process which we survey on a frame by frame basis:
the \mpdrei format encodes an audio stream into a series of \mpdrei frames. 
Each of these frames contains a header, containing format information and  parameters of the \mpdrei encoding process itself. 
An optional integrity check for the header is often omitted to save bitrate. 
The \texttt{lame}\footnote{\url{https://linux.die.net/man/1/lame}} encoder used in the present study does not include CRC checking.
The header is followed by the encoded audio data, the size of this block being determined by the bitrate used for the frame during encoding. 
If a corrupted file remains undetected before and during decoding, introduced errors may be audible as glitches in playback, but become (from a data integrity point of view) undistinguishable from the original signal. 
This may easily happen if a command-line decoding tool only issues a warning in case of a data corruption which the decoding process can recover from in successive frames.

We generate glitches as follows: 
First, the input audio is transcoded to 128kbps mono \mpdrei files.
During the following decoding, frames are randomly selected for glitch insertion given a probability of $P_{glitch}=0.05$. 
In contrast to click generation we control glitch likelihood per \mpdrei frame. 
The data in these frames is then partially overwritten with random data of a randomised length. 
We found an average overwritten range of 120 bytes (an average frame contains 418 bytes) %
with a standard deviation of 60 bytes to give realistic results.
No distinction was made between the header and data sections of the \mpdrei frame.
In the rare case that the decoding of the degraded frame is not possible, the original \mpdrei frame is decoded and treated as non-degenerated. 
The same method is used where the decoded data is of a different length than the original un-degraded signal.
Frames not selected for glitch insertion are decoded inbetween such that the degraded decoding of the \mpdrei data results in a file with the same length as the original.

\subsection{\mpdrei glitch target vector}
\mpdrei glitches are inserted during pre-processing.
To identify the actual parts of the decoded wave signal affected by the glitch artefact after decoding and segmentation, we employ a spectral distance measure comparing the degradedly decoded audio to a clean decoding of the audio.
Using a frame wise thresholded difference of a power spectrogram at the frame rate of the network output (128 spectrogram frames), we determine whether significant degradation has taken place for each of the 128 output values.
Fig. \ref{fig:glitch_hf} (bottom) shows a resulting glitch classification target.

\section{Experiments}
We compare our detectors on the Creative-Commons licensed FMA-Large\footnote{Online at \url{https://github.com/mdeff/fma/}.} dataset, containing roughly 30-second snippets of 106,574 tracks from 16341 artists within 161 genres, the most frequent being "Experimental", "Electronic", and "Rock"\cite{fma_dataset}.
The music is stored in 320kbps stereo \mpdrei format.

Depending on the scenario, the data is either decompressed and degraded on-the-fly, or, as in the \mpdrei glitch scenario, data is already pre-processed with degradations added, and loaded as raw waveform alongside corresponding ground truth data.
The \hooknet models take as input audio segments of 16384 samples at a sampling rate of 44.1kHz, corresponding to 0.37 seconds. 

Degraded (as well as non-degraded) audio and target data are then used for network training, using a batch size of 200 segments. 
During configuration of the \hooknet, we tested variations on general parameters such as numbers of filters and found the models with 13 contracting/expanding blocks (see \Cref{sec:flatunet}), 15 filters per contraction and 5 filters per expansion, totalling at 27,307,633 trainable parameters, performing well for our tasks at hand.
Initial experiments consistently showed reduced precision in smaller models. 
Training is performed in Tensorflow, using the ADAM\cite{kingma2014adam} optimiser, on single Nvidia V100 GPU.
\subsection{Click detection}\label{sec:experimentsclick}\label{sec:datasetclick}
We selected a subset of 57,928 pieces from FMA-Large for training and evaluation of our click detector \hooknet.
Pieces were separated into non-overlapping training (40,000 
pieces), validation (8964 pieces), and test (8964 pieces) sets. 
For each piece, 50 consecutive audio segments (18.6s in total) were extracted with no overlap, resulting in 2,000,000 training segments and 448,200 validation and 448,200 test segments. 
In the above datasets, 0.078\% of the individual target values is set to 1 (on average, in every 10th segment, one value in the target vector is marked as containing a click), which is reflected in the initial setting of the network outputs' activation biases.
Evaluation is performed on the target values as smallest units,  not summarised at the segment level.
The model was trained using a root mean square loss weighted towards the ``click'' class with a learning rate of 0.001. 
We report results from models of the epoch with best validation set accuracy. 
 
As a baseline for the click detection task we apply a generic click detector  (\esclick) 
as implemented\footnote{\url{https://essentia.upf.edu/reference/std_ClickDetector.html}} in the open source Essentia library\cite{Bodganov2013Essentia} to the data segments.  
To apply this to our segments of 16384 samples, we add an additional sub-segmentation step, using sub-segments of 4096 samples length, with an overlap of 2048 samples.
Click positions returned by \esclick are transformed into a binary output vector of 128 samples. 

\begin{table}
		\small 
		\begin{tabular}{|c|c|c|c|c|c|c|} 
			\hline
			 & test data & acc\_t &    pr\_t   &  rec\_t  & f1\_t    \\ 
			\hline
			\mfull & FMA & \textbf{99.9993} & \textbf{99.77} & \textbf{99.24} & \textbf{99.47} \\ 
			\esclick  & FMA & 99.95 &  84.39 & 90.52 &  84.60 \\ 
			\hline
			\mfull & $\text{FMA}_{\text{post}}$ & \textbf{99.9991} & \textbf{99.70} & \textbf{99.02} & \textbf{99.32} \\
			$\text{\mfull}_{\text{post}}$ & $\text{FMA}_{\text{post}}$ & \textbf{99.9995} & \textbf{99.86} & \textbf{99.11} & \textbf{99.46} \\
		\esclick &$\text{FMA}_{\text{post}}$ & 99.97 &  85.52 & 81.04 &  80.62 \\ 
			\hline
\end{tabular}
\caption{Click detector performances for plain (top) and post-processed (bottom) click-degraded data (higher is better, best per dataset in bold): test-set accuracy (acc\_t), test precision(pr\_t), recall(rec\_t) and f-measure (f1\_t) for click detection. In percent. \label{tbl:clickperf}
}
\end{table}
\Cref{tbl:clickperf} compares the test set performance of the \hooknet to that of the best performing (by highest validation set accuracy) configuration (threshold 35) of the \esclick detector between tested thresholds of 30(default), 33, 35, 40, and 50.
The \hooknet models were selected by highest validation set accuracy, which was achieved after 13 epochs (36 total) for the click data (\mfull) and 10 epochs (14 total) for training with post-processed click data ($\text{\mfull}_{\text{post}}$).

Due to the bias of the dataset and target vectors towards 0 (no click), we concentrate on precision and recall which is reported for the click class.
For the application of finding defects in large commercial databases, the precision of click detection is of great importance.
Top of \Cref{tbl:clickperf} shows the \hooknet achieves significantly better precision and recall of clicks. 
The bottom of the table confirms this using post-processed click data as test set.
Added variation from post-processing results in lower recall values, particularly with \esclick.
The difference in training with (\mfull$_{\text{post}}$) and without (\mfull) post-processing shows good generalisation from the generated click artefacts to the diverse post-processed artefacts.

These results highlight a critical difference in the data-driven paradigm applied in network training versus the more generalistic detection in \esclick: while the assumption for a general click detector is to detect any existing click (given other preconditions e.g. sufficient distance to the preceding one), the \hooknet has been trained to detect our inserted clicks while ignoring other click instances in the music.
Note that we do only compare the performances on our clicks simulating digital defects such as buffer under-run and timing errors. 
The \esclick detector may detect a wider range of clicks, but for this initial experiment we refrained from training more generic models due to the lack of datasets with clicks stemming from defects in music production, as we aim to minimise false positives.

Manual verification showed that the FMA dataset does feature many examples of electronic ``glitch'' music with intentional or expected clicks. 
The goal of our detector is to differentiate such clicks from the ones added due to signal failures.

\begin{table}
        \small 
        \begin{tabular}{|c|c|c|c|c|c|c|} 
            \hline
             & dataset &  acc\_t &    pr\_t   &  rec\_t  & f1\_t \\ 
            \hline
            \mfull & electronic & \textbf{99.9997} & \textbf{99.91} & \textbf{99.74} & \textbf{99.82} \\ 
            \esclick &  electronic & 99.94 & 80.34 & 98.06 &  84.94 \\ 
             \hline
            \mfull & rock pop & \textbf{99.99992} & \textbf{100.0} & \textbf{99.90} & \textbf{99.95} \\ 
            \esclick & rock pop & 99.99 &  92.08 & 94.37 &  92.05 \\ 
            \hline
            \mfull & classical & \textbf{100.0} & \textbf{100.0} & \textbf{100.0} & \textbf{100.0} \\ 
            \esclick & classical & 99.992 &  92.47 & 99.11 &  95.34 \\
            \hline
\end{tabular}
 \caption{Click performance in control datasets. Test-set accuracy (acc\_t), test precision(pr\_t), recall(rec\_t) and f-measure (f1\_t).\label{tbl:clickperfclassic}  
 }
\end{table}

We validate this relation to musical genre in \Cref{tbl:clickperfclassic} by applying the above model and \esclick on three smaller genre-consistent datasets of 11400 (electronic), 10000 (pop rock) and 31600 (classical) segments not included in the former training.
Following the hypothesis that the electronic genre features more intentional click samples than the classical genre, we see the precision of \esclick very high for the classical data but significantly dropping in the rock/pop and electronic genres, while the \hooknet maintains high precision with only little difference.
The values for post-processed data not reported here due to space limitations confirm this effect.

\subsection{\mpdrei glitch detection}\label{sec:experiments}\label{sec:dataset}
For this task, pre-caching of degraded audio allows us to use  larger subsets of FMA for training (66476 pieces), validation (14244 pieces) and test (14244 pieces). 
	
For each piece, 50 consecutive audio segments were extracted as above, resulting in 3,323,800 training segments and 712,200 validation and test segments. 
Segments were randomised within each of the above datasets. 
While a click only results in 1 target value to be set to 1, glitches affect multiple target values per segment due to whole \mpdrei frames being affected by each corruption. 
This results in train and validation datasets containing 8.04\% of the target values marked as glitched (test set: 8.19\%). 

Training is performed using root mean square loss, with initial learning rate of 0.001 and reduction-on-plateau (factor 0.1, patience 10 epochs) of the learning rate. 
Across 40 training epochs, the best model was selected based on its validation accuracy (acc\_v) measure.
\begin{table}
	\small 
	\begin{tabular}{|c|c|c|c|c|c|c|} 
		\hline
		& ep. & acc\_v & acc\_t &    pr\_t   &  rec\_t  & f1\_t    \\ 
		\hline
		\mfull  & 31    & \textbf{98.52}  & \textbf{ 98.46} & 92.83 &\textbf{ 88.04} & \textbf{90.37} \\ 
		\hline
	\end{tabular}
	\caption{Glitch detection performance. Epoch with greatest validation-set accuracy, validation accuracy (acc\_v), test-set accuracy (acc\_t), test precision(pr\_t), recall(rec\_t) and f-measure (f1\_t) for glitch detection. In percent.\label{tbl:compperf} 
	}
\end{table}
\Cref{tbl:compperf} shows the performance of this best model. 
Given the bias towards the non-degraded class we report f-measure and precision regarding the glitched class due to their relevance for our application scenario.
The \mfull glitch model generalises well from the validation to the test set, with an f1-measure of 0.9037.

\section{Conclusions and Future Work}\label{sec:conclusions}
We presented the \hooknet convolutional neural network architecture with two novel applications to detect click and data corruption errors in digital audio recordings.
The design goal of the architecture is to capture the typical shape and variation of artefacts in the direct audio signal, with respect to their acoustic context.
The detectors localise errors with a time resolution of less than 10 milliseconds.

For click detection, we demonstrated an end-to-end simulation, post-processing and detector training method.
Our evaluation shows the resulting detector outperforms a state-of-the-art baseline on the large FMA popular music dataset, using synthetically generated defects. 
A genre-specific evaluation experiment shows the practical relevance of inclusion of context and the capability of our model to capture this: 
discontinuities resembling clicks in audio may represent intentional music content depending on their context.
The achieved precision ($>99.77$\%) renders the detector suitable for the testing of large commercial music databases.

Our second proposed application is aimed to filter degradations in modern compressed audio
from persisting unnoticed in subsequent use of the defect audio.
We describe the \mpdrei decoding glitch as a relatively novel type of audio degradation for which detection is difficult due to its variation. 
Our evaluation shows that our detector generalises well on common glitch artefacts. 
The proposed training tasks come with a bias towards non-defective audio, which we assume to be strong in real-world applications. 
This is tackled using large training and validation datasets with synthesised artefact insertion.
The simulation of data corruption with subsequential \mpdrei decoding promises a stronger realism of the artefacts synthesized in this task. 
This process is applicable to other audio codecs, depending on decoder robustness and consistency checks.

In future work we plan to extend the range of lossy compression defects simulated and apply our architecture to further and more generalised local audio degradations. 
We also aim to reduce model complexity without negatively affecting performance.
\section{Acknowledgements}
This work was performed using HPC resources from GENCI-IDRIS (Grant 2020-AD011011379).

\bibliography{refs}

\end{document}